\documentclass[aps,twocolumn,prl]{revtex4}
\usepackage{graphicx}
\usepackage{amsfonts}
\usepackage{amssymb}
\usepackage{amsmath}

\begin{document}                                                                    
    
\title{Non-Nearest-Neighbor Interactions in
 Nonlinear Dynamical Lattices}
\author{
P. G.\ Kevrekidis}
\affiliation{
Department of Mathematics and Statistics, University of Massachusetts, Amherst MA 01003-4515, USA}

\begin{abstract} 
We revisit the theme of non-nearest-neighbor interactions in nonlinear
dynamical lattices, in the prototypical setting of the discrete nonlinear
Schr{\"o}dinger equation. Our approach offers a systematic way of
analyzing the existence and stability of solutions of 
the system near the so-called anti-continuum limit of zero
coupling. This affords us a number of analytical insights such as
the fact that, for instance, next-nearest-neighbor interactions
allow for solutions with nontrivial phase structure in infinite 
one-dimensional
lattices; in the case of purely nearest-neighbor interactions, such
phase structure is disallowed. On the other hand, 
such non-nearest-neighbor interactions can critically
affect the stability of unstable structures, such as topological charge
$S=2$ discrete vortices. These analytical predictions are corroborated
by numerical bifurcation and stability computations.
\end{abstract}

\maketitle

\section{Introduction}

Undoubtedly, the discrete nonlinear
Schr{\"o}dinger (DNLS) equation is one of the most
fundamental nonlinear lattice dynamical models.
Part of its relevance is due to its being one of the
main discrete analogs of the famous (and integrable in
1+1 dimensions) continuum nonlinear Schr{\"o}dinger (NLS)
equation \cite{sulem,trub}. The latter is encountered in 
a wide
spectrum of applications; it is the relevant dispersive envelope wave
model for the electric field in optical fibers
\cite{hasegawa,malomed}, for the self-focusing and collapse
of Langmuir waves in plasma physics \cite{zakh1,zakh2}
or for the description of freak waves (the so-called rogue waves)
in the ocean \cite{onofrio}.

On the other hand, the DNLS model is of physical interest
in its own right within a diverse host of applications; see
the reviews \cite{IJMPB,joh_rev}, as well as the recent book \cite{kev_book}.
One of the principal directions that spurred a considerable
interest in the DNLS model was the area of the nonlinear
optics of fabricated AlGaAs waveguide arrays \cite{7}. 
In this seting, a plethora of important phenomena have been experimentally
investigated including discrete diffraction, Peierls barriers
(the energetic barrier that a wave needs to overcome to move
over a lattice),
diffraction management (the periodic alternation of the
diffraction coefficient) \cite{7a} and gap solitons 
(structures localized due to nonlinearity in the gap of the underlying 
linear spectrum) \cite{7b} among
others \cite{eis3}. More recently, quasi-periodic \cite{recent1}
and even completely disordered \cite{recent2} variants of such
waveguide lattices have been constructed, allowing the observation of
localization type phenomena (even within the linear regime for
ones of the Anderson type). Many of these results and the 
corresponding theoretical activity that they triggered has been
summarized in a number of reviews 
see e.g. \cite{review_opt,general_review}.

Another independent and considerably different physical setting
where DNLS-type models naturally arise is that of Bose-Einstein
condensates (BECs) trapped within periodic (so-called, optical
lattice) potentials created by counter-propagating laser beams
in one-, two- or even three- spatial dimensions \cite{bloch}.
This area of atomic physics has  also experienced a huge growth over 
the past few years, including the prediction and manifestation of 
modulational instabilities
(i.e., the instability of spatially uniform states towards spatially 
modulated
ones)
\cite{pgk}, the observation of gap solitons \cite{markus}, Landau-Zener
tunneling (tunneling between different bands of the periodic potential) 
\cite{arimondo} and Bloch oscillations (for matter waves subject
to combined periodic and linear potentials) \cite{bpa_kasevich} 
among many other salient features; reviews of
the theoretical and experimental findings in this area have also 
recently appeared in \cite{konotop,markus2}.

Our aim in the present communication is to consider a somewhat non-standard
variant of the DNLS equation, namely one in which the interaction
kernel within the linear term allows interactions among {\it any}
set of neighbors. Such, so-called, nonlocal versions of the DNLS
model have been abundantly considered in the past, yielding
numerous interesting features. In particular, it was shown, for
instance,
that, if the interaction strength decays sufficiently slowly as a function
of distance, it can give rise to bistability of fundamental solitary
waves (centered on a single site) \cite
{nonlocalBistability}, which may find applications in their controllable
switching \cite{switchingBistable}. Dynamical lattices with long-range
interactions also serve as models for energy and charge transport in
biological molecules \cite{EnergyChargeTransportBiomolecules}. Systems
including the next-nearest-neighbor (NNN) interaction have also been
proposed as models of polymers \cite{nextnearestPolymer}, as well
as in the case of optical waveguide arrays near the
zero-dispersion point \cite{borisus}. Nonlinear lattice
models with competing short- and long-range interactions were studied not
only in one, but also in two dimensions \cite{2DcompetingShortLongRange}.
Lastly, it is relevant to mention that quantum DNLS models with nonlocal
interactions were another subject of study (by means of the Bethe
ansatz) \cite{quantum}. It should be pointed out that
such longer-range interactions are often directly relevant to physical 
applications. For example, in the case of the waveguide arrays,
the NNN interaction is also, in principle, present yet
its relative strength depends on the ratio of the separation between
adjacent cores to the wavelength of light. A specific, zigzag-shaped, model
of an optical array involving a NNN coupling was introduced in
Ref. \cite{Cyprus}, allowing for essentially arbitrary ratios
of nearest- to next-nearest-neighbor interactions, depending on the
specific geometry. Lately, long-range interactions, especially
of a dipolar type have become of considerable interest in BECs due
to the significant component of dipole-dipole interactions in
$^{52}$Cr. These effects, as well as the significant component of
studies of such dipolar BECs in optical lattices have recently been
summarized in \cite{pfau}.

The scope of the present communication is to explore the non-local
variant of DNLS in the vicinity of the so-called anti-continuum (AC)
limit, where all individual sites (i.e., waveguides in optics or
wells of the optical lattice in BEC) are uncoupled. From there,
in the spirit of \cite{peli1d}, we can develop a
perturbative
expansion using the strength of the interactions as the relevant
small parameter. This allows us to derive 
conditions for the persistence of states near the AC-limit,
as well as to formulate a framework to address their linear stability.
Both the existence conditions, as well as the stability conditions
allow us to infer valuable conclusions. The existence 
conditions suggest (and numerical computations corroborate)
the possibility that solutions with a {\it nontrivial phase distribution}
may arise in the 1d setting, a feature which is absent in the
case of nearest-neighbor (NN) interactions; even the sole inclusion
of NNN interactions allows to achieve this feature. On the
other hand, concerning stability, the inclusion of NNN interactions
may also play an important role, as we illustrate with a 
two-dimensional example: the stabilization of a topological
charge $S=2$ vortex (which is unstable in the presence of
NN interactions) is numerically observed and
analytically justified.

Our analytical considerations are presented in section II, while
section III is devoted to the corresponding numerical results (and
their comparison to theory). Finally, section IV summarizes our
findings
and
suggests a number of interesting future directions.

\section{Analytical Considerations}

Our nonlocal variant of the DNLS equation will be of the
following form:
\begin{eqnarray}
i \dot{u}_n - u_n + |u_n|^2 u_n = -\varepsilon \sum_{m=1}^N k_{nm} u_m
\label{lrdnls}
\end{eqnarray}
where the overdot denotes temporal (in the case of BECs in optical
lattices) or spatial (in the case of waveguide arrays) derivatives,
$n$ denotes the site index (waveguide index in the optical case,
or optical lattice well in BECs),
and $\epsilon$ controls the strength of the coupling among sites, 
which is also
determined by the ``coupling matrix'' with elements $k_{nm}$.
The dependent variable $u$ represents the complex 
envelope of the electric field in optics, or the complex wavefunction
of the BEC in atomic physics.
The AC limit is defined by $\varepsilon=0$, in which case the
unperturbed energy of the uncoupled oscillators reads
\begin{eqnarray}
E_0(u)=\sum_n |u_n|^2 - \frac{1}{2} |u_n|^4.
\label{lrdnls2}
\end{eqnarray}
Once the inter-site coupling is turned on, the relevant perturbation
in the
energy
reads: 
\begin{eqnarray}
E_1(u)=-\epsilon \sum_{n,m=1}^N k_{nm} \left(u_n^{\star} u_m + u_n u_m^{\star}
\right)
\label{lrdnls3}
\end{eqnarray}
Notice that at the AC limit,  the solutions for the excited
oscillators will be $u_n=e^{i \theta_n}$, where the $\theta_n$'s
are arbitrary phase parameters ($u_n=0$ corresponds to the non-excited
sites).

To determine the persistence of the waves, as was illustrated
in \cite{todd_pers} (see also the general framework of
\cite{bjorn} or the nearest-neighbor case of \cite{peli1d})
one has to evaluate
the perturbed energy at the unperturbed limit solution, in which
case, we can straightforwardly evaluate it to be:
\begin{eqnarray}
E_1=-\sum_{n,m=1}^N 2 k_{nm} \cos(\theta_n-\theta_m),
\label{lrdnls4}
\end{eqnarray}
where $N$ is the number of excited (adjacent in this case,
for simplicity, --although the theory can be appropriately
generalized for non-adjacent) oscillators.
The general persistence conditions then read $\partial_{\theta_n} E_1=0$ 
\cite{todd_pers,bjorn}, i.e., the unperturbed wave needs to correspond to an
extremum of the perturbation energy in order to persist.
This leads to the condition
\begin{eqnarray}
\sum_{m \neq n} k_{nm} \sin(\theta_n-\theta_m)=0.
\label{lrdnls5}
\end{eqnarray}

On the other hand, the general stability theory of 
\cite{bjorn} (again see \cite{peli1d} for a particular
application in the case of nearest-neighbor interactions)
allows us to quantify the bifurcation of the $N-1$ eigenvalues
from $\lambda=0$. There are $N$ eigenvalues at
$\lambda=0$ when $\varepsilon=0$, and only one of them remains
there for finite $\varepsilon$, 
due to the U$(1)$ symmetry and associated phase invariance,
while $N-1$ acquire small (O$(\sqrt{\varepsilon})$ or smaller)
values, as the sites become coupled.  These eigenvalues
satisfy the reduced eigenvalue problem \cite{bjorn}
\begin{eqnarray}
(D_0 \lambda^2 + M) v=0,
\label{lrdnls5a}
\end{eqnarray}
where
\begin{eqnarray}
D_0 &=& - I_N
\label{lrdnls6}
\\
M_{nm} &=& \partial_{\theta_n \theta_m}^2 E_1,
\label{lrdnls7}
\end{eqnarray}
i.e., the stability is effectively determined by the
{\it Jacobian} of the persistence conditions.
It is worthwhile to note that this general formulation 
encompasses the nearest-neighbor one of \cite{peli1d}
as a special case. 

The above formulas, namely Eq. (\ref{lrdnls5}) describing
the existence and Eqs. (\ref{lrdnls5a})-(\ref{lrdnls7})
quantifying the stability under weak coupling are the main
analytical results of the present study, which we now proceed
to utilize in specific applications where additional
(to the nearest neighbor) interactions exist.
 
\section{Numerical Results}

In this section, we apply the above analysis to a few select
cases where interactions other than purely nearest neighbor
ones are included, so as to provide a sense of the kind of
additional features that such settings may entail and how
these features are brought forth through the above analysis.
We start with what arguably can be dubbed the simplest extension
of NN interactions, namely the inclusion of the NNN ones which is
directly physically realizable, as per the setting of \cite{Cyprus}.

\begin{figure}[tbp]
\includegraphics[width=4cm,height=5cm,angle=0,clip]{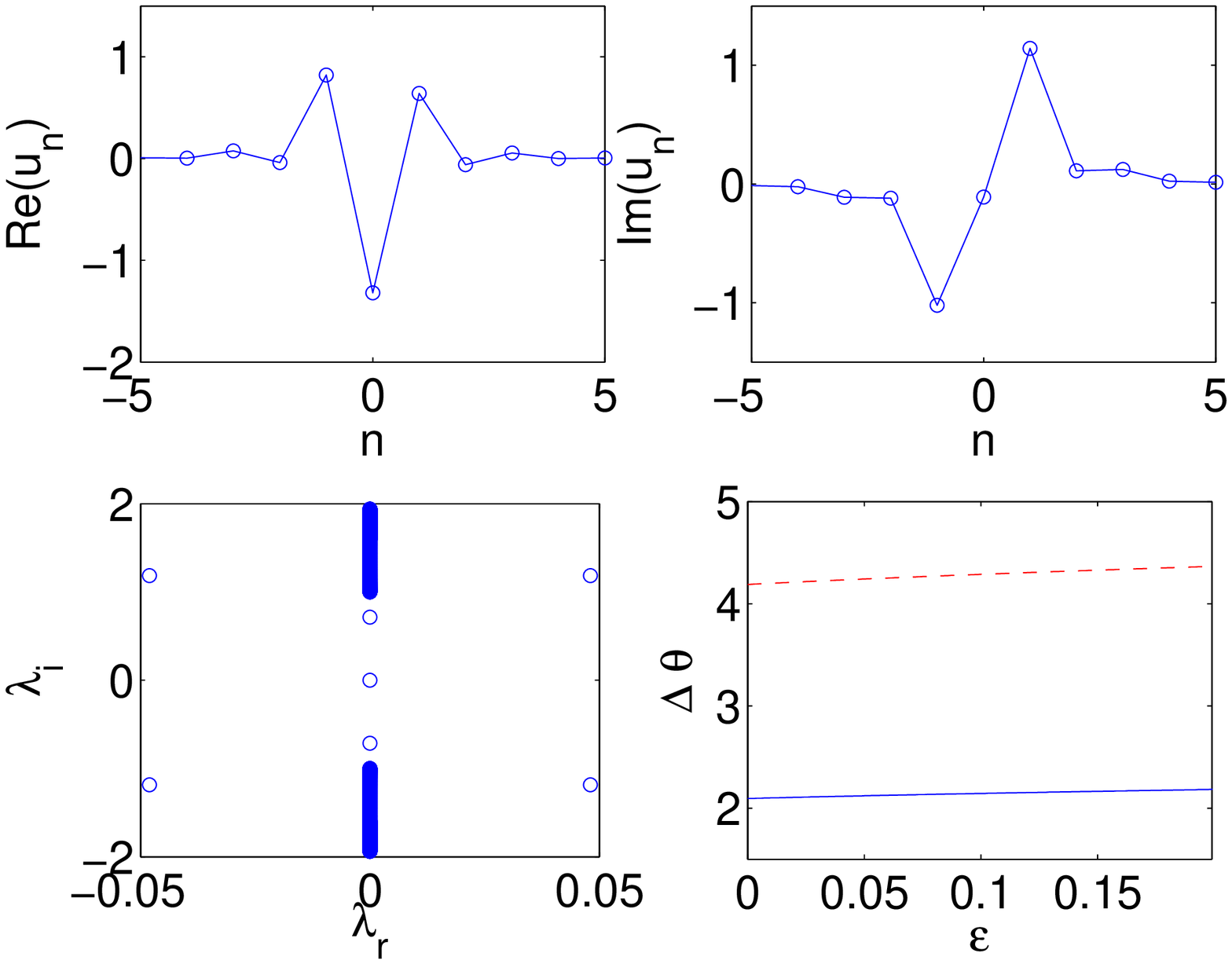}
\includegraphics[width=4cm,height=5cm,angle=0,clip]{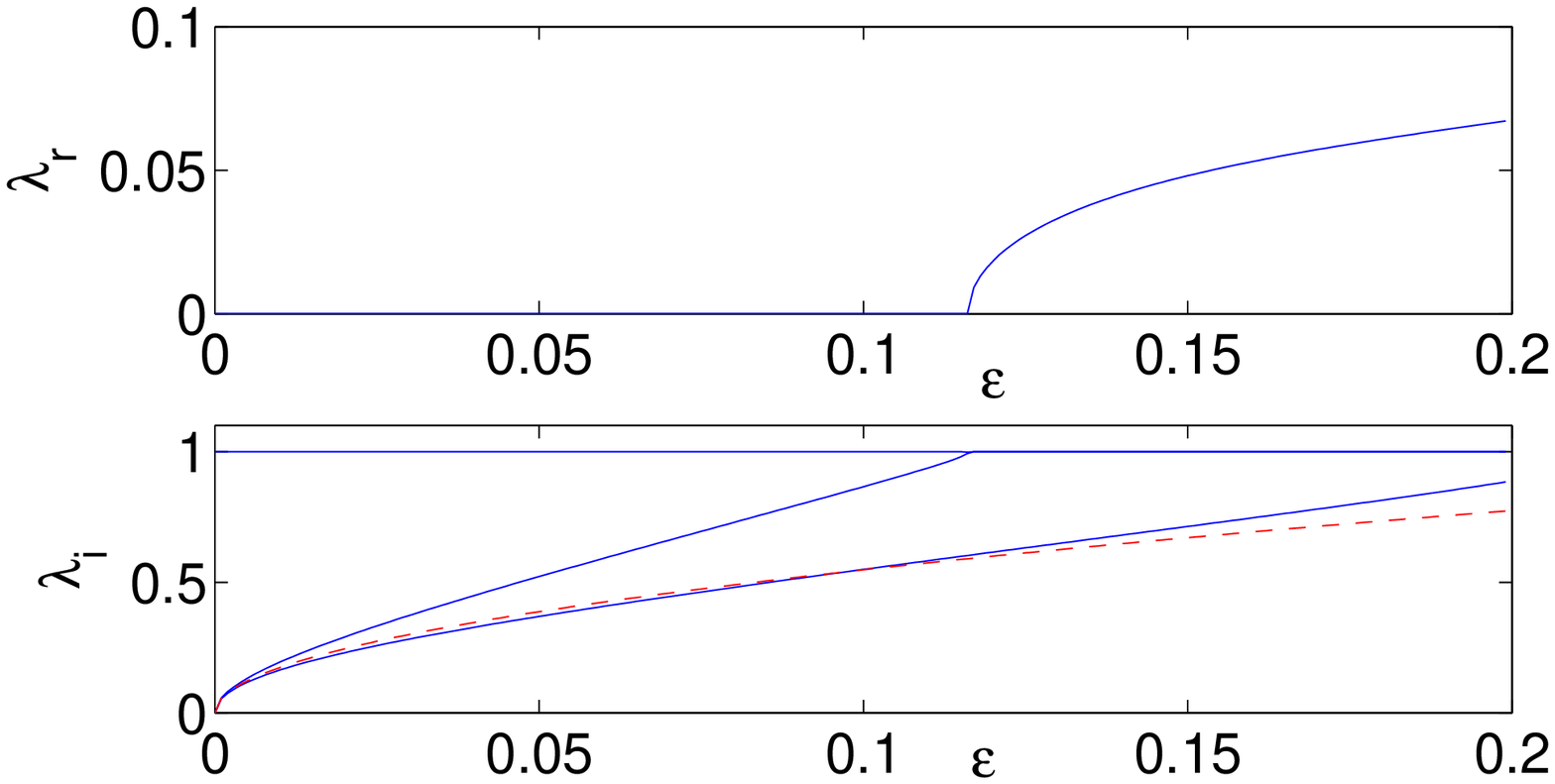}
\caption{The figure shows a 3-site 
solution with a nontrivial phase distribution
$(0,2 \pi/3,4 \pi/3)$ in the 1d case with equal nearest and
next-nearest-neighbor
interaction strengths of unity, i.e., $k_{n,n \pm 2}=k_{n,n \pm 1}=1$. 
The first two panels
in the top row show the real and imaginary part of the solution for
$\varepsilon=0.15$, while the bottom left shows the corresponding
spectral plane $(\lambda_r,\lambda_i)$ of the linearization eigenvalues
$\lambda=\lambda_r+i \lambda_i$. The second panel of the bottom
row shows the relative phases $\Delta \theta=\theta_2-\theta_1$ (blue
solid line) and $\theta_3-\theta_1$ (red dashed line), which does
not change significantly from its anti-continuum limit as
$\varepsilon$ is increased. Lastly, the top right and bottom rightmost
panels show the maximal real eigenvalue (larger than zero -and hence
giving rise to an instability- 
for $\varepsilon > 0.116$) and lowest imaginary eigenvalues,
respectively (blue solid lines). The theoretical prediction 
for the latter is given
by the red dashed line.} 
\label{fig1}
\end{figure}

We start from the case of equal NN and NNN interactions only, i.e.,
$k_{nm}=\delta_{m,n \pm 1} + \delta_{m,n \pm 2}$, where $\delta$ denotes
the Kronecker $\delta$. Considering then a $N=3$ site configuration
(the smallest one that ``perceives'' of the presence of the NNN
interaction), it is straightforward to intuitively realize that
this is essentially the analog of an equilateral triangle where
all pairwise interactions exist and are equal to each other.
This is a situation that has been considered both in DNLS
\cite{kody_pre} (see also the early work of \cite{chris_e})
and in Klein-Gordon \cite{vass1,vass2} type settings, 
motivated by applications of triangular lattices in 
photorefractive crystals and dusty plasmas. It has been
illustrated therein that the fundamental {\it linearly stable}
3-excited-site configuration is that of a so-called vortex of topological charge
$S=1$, namely a situation where the phase runs from $0$ at the
first excited site, to $2 \pi/3$ at the second one and $4 \pi/3$
at the third one. Remarkably, all configurations with
the standard (from the NN setting of \cite{peli1d}) case of
$0$ or $\pi$ phases, are unstable.
Equally importantly the NN case, as per the arguments of
\cite{peli1d},
never allows for the existence of phases other than $0$ or $\pi$
in 1d configurations. Hence, one of the principal features of
these non-nearest-neighbor interactions is that they may enable
configurations that would not be possible in standard NN settings.
Moreover, as indicated above in this case, the eigenvalues can
be analytically computed (through a discrete Fourier transform applied
to the ``triangle'' of sites (which has periodic boundary
conditions). This can be found to lead to the eigenvalues
\begin{eqnarray}
\lambda_j= \pm \sqrt{8 \varepsilon \cos(\Delta \theta) \sin^2
\left(\frac{\pi j}{3}\right)}
\label{lrdnls8}
\end{eqnarray}
for $j=1,2,3$. The stable case features the relative phase
between adjacent sites $\Delta \theta=2 \pi/3$ and yields a
double pair of imaginary eigenvalues $\pm \sqrt{3 \varepsilon} i$
which are compared to the full numerical results in Fig. \ref{fig1}
(also the analytical existence prediction for the relative angle among
adjacent sites is tested through the numerical
results of the figure). Good agreement
is obtained between the analytical and numerical results in both
existence and stability, for small values of $\varepsilon$,
while, as expected, relevant deviations increase for larger values
of the perturbation parameter. We also note in passing that the
non-trivial phase vortex structure becomes unstable eventually in
an oscillatory way
due to a complex quartet of eigenvalues for $\varepsilon > 0.116$
as also illustrated in Fig. \ref{fig1}, due to the collision of
one of the imaginary pairs with the continuous spectrum extending
above $\lambda=i$ (and below $\lambda=-i$).

\begin{figure}[tbp]
\includegraphics[width=4cm,height=5cm,angle=0,clip]{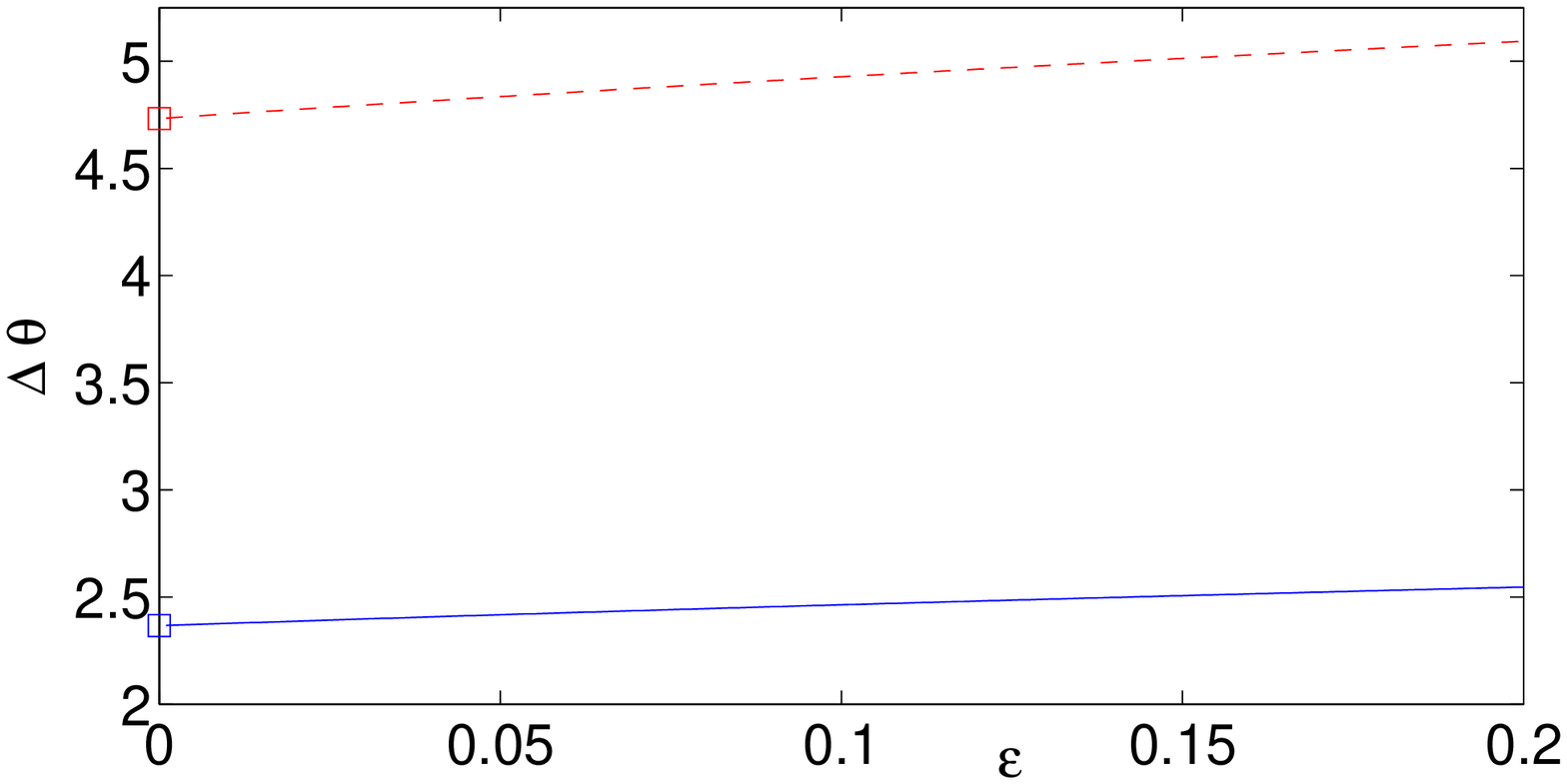}
\includegraphics[width=4cm,height=5cm,angle=0,clip]{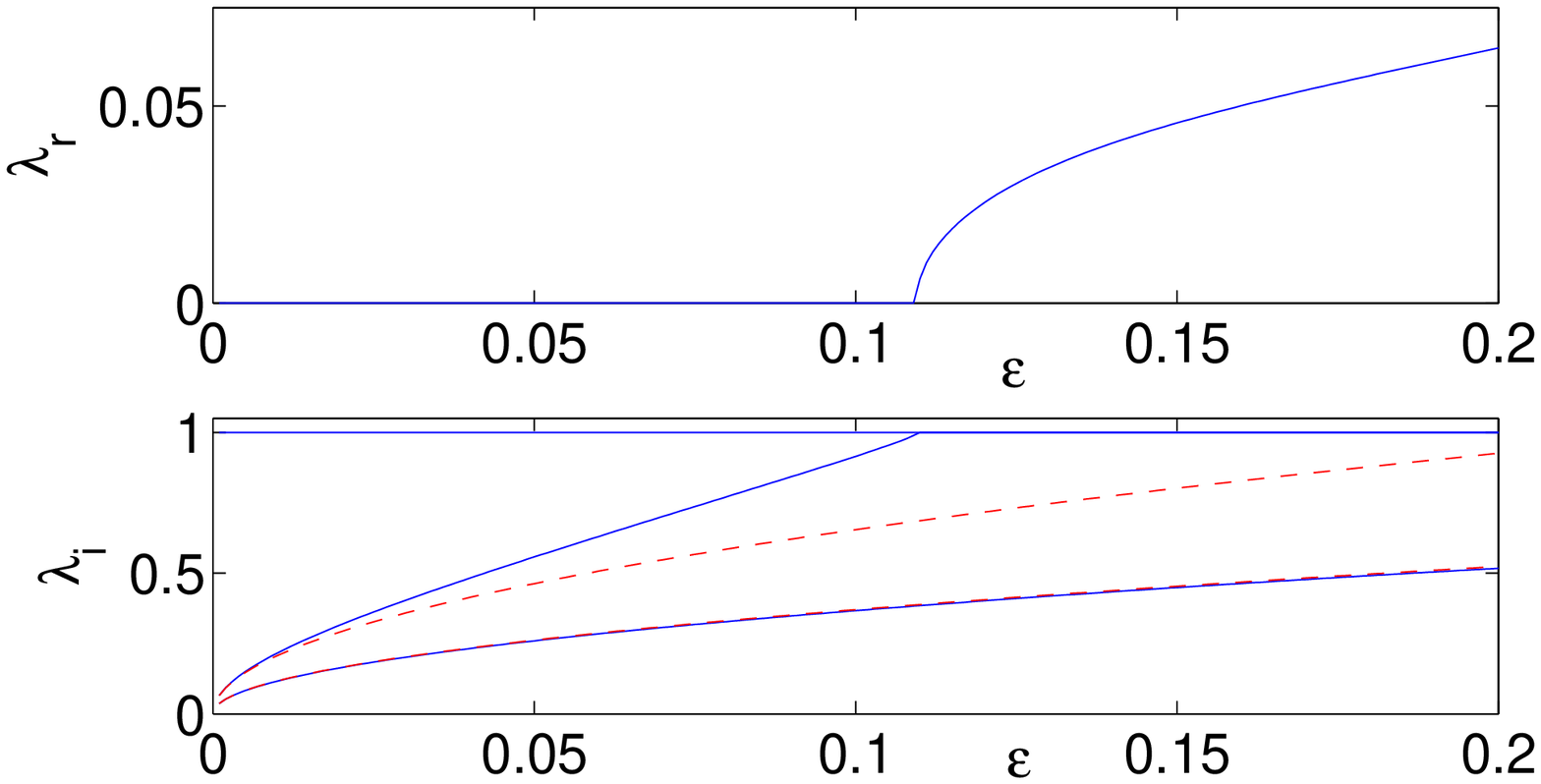}
\caption{An example of weaker (than nearest-neighbor)
next-nearest-neighbor interaction strength with $k_{n,n \pm 2}=0.7$ while 
$k_{n,n \pm 1}=1$:
the phase distribution is shown on the left (and its anti-continuum
limit, as theoretically predicted is depicted by squares), while
the predictions of the linear stability analysis (red dashed lines)
are compared with the full numerical results (blue solid lines) 
as a function of  $\varepsilon$ on the right. The real part of the
corresponding eigenvalues becomes nonzero for $\varepsilon > 0.109$
(as shown in the top right panel).} 
\label{fig2}
\end{figure}

To illustrate the generality of our analytical considerations
in settings other than the special, degenerate case of
Fig. \ref{fig1},
we considered another physically realizable (as per \cite{Cyprus})
case where $k_{n,n \pm 2}=0.7 k_{n,n \pm 1}=0.7$ (while all other interactions
are absent). Again, focusing on the 3-site interaction, we realize
that the three persistence conditions can be explicitly solved,
yielding two possible scenarios. Either 
$\sin(\theta_n-\theta_m)=0$
for all combinations of $n,m \in \{1,2,3\}$ {\it or}
$\theta_1-\theta_2=\theta_2-\theta_3$ and 
$\cos(\theta_1-\theta_2)=-k_{n,n+1}/(2 k_{n,n+2})$. 
Notice that this encompasses the discussion of
equal strengths of NN and NNN interactions as a special
case and additionally reveals the condition under which
the nontrivial phase distributions discussed above 
can generally materialize, since it is necessary that
$|k_{n,n+1}/(2 k_{n,n+2})|<1$. In the case of Fig. \ref{fig2},
the analytical prediction yields a phase difference
of $\theta_1-\theta_2=2.3664$ ($\theta_1-\theta_3$ is
twice that) in excellent agreement with the numerical
observations. Notice also that the relative phases
appear to depend only  weakly on $\varepsilon$
in the continuations over the latter parameter shown in
Fig. \ref{fig2}. Additionally, our stability considerations
allow us in this case to obtain the analytical predictions
for the two (now split due to the asymmetry in the 
interactions) pairs of small eigenvalues 
$\lambda=\pm \sqrt{4.2858 \varepsilon} i$
and $\lambda=\pm \sqrt{1.3714 \varepsilon} i$. Good agreement
is once again found with the full numerical linear stability
results for small couplings $\varepsilon$, while this
agreement deteriorates when increasing the coupling. The
solution becomes unstable once again through the oscillatory
instability arising upon collision (for
$\varepsilon>0.109$) of the larger one among the above 2
pairs with the continuous spectrum.

\begin{figure}[tbp]
\includegraphics[width=4cm,height=5cm,angle=0,clip]{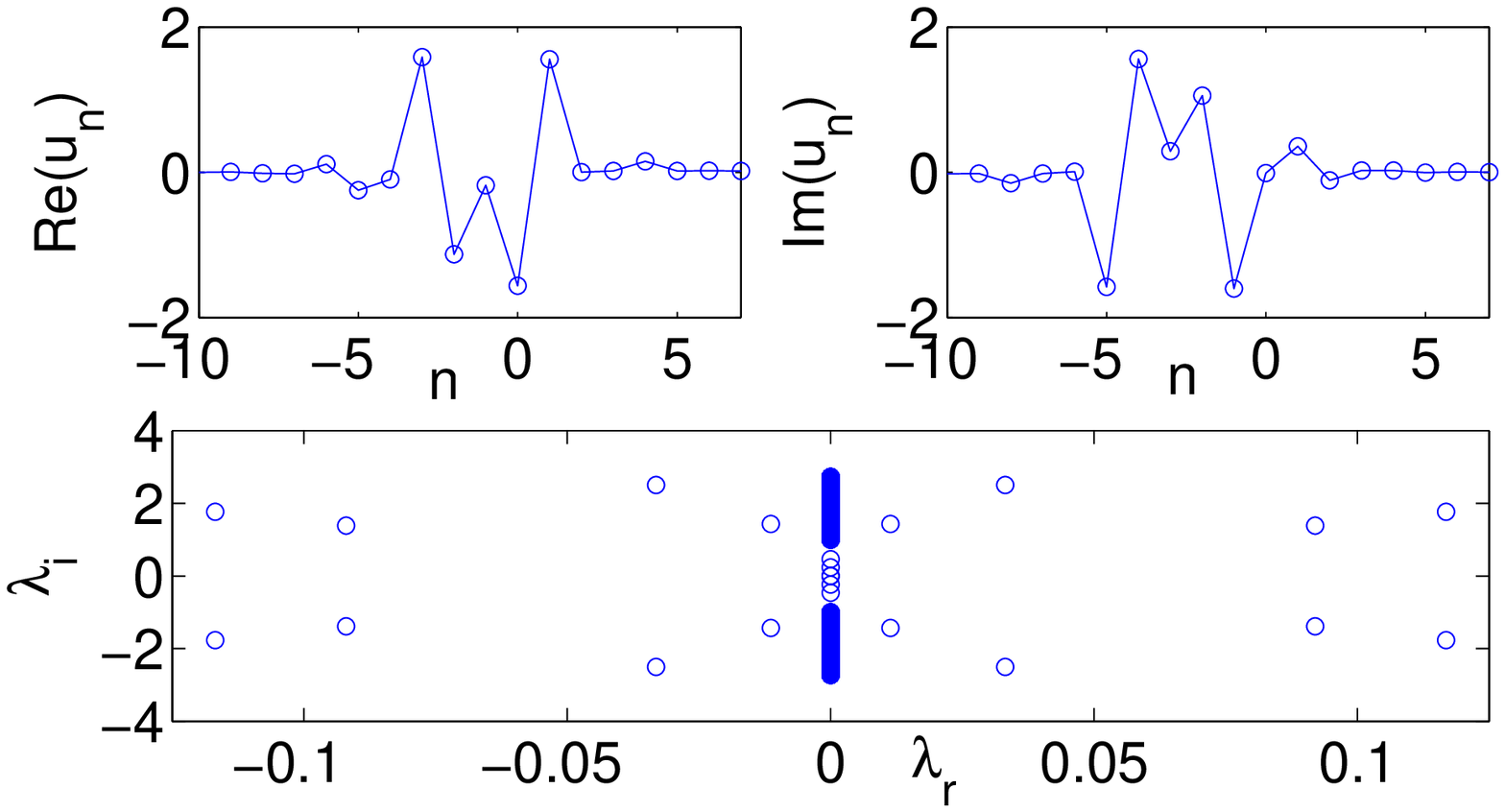}
\includegraphics[width=4cm,height=5cm,angle=0,clip]{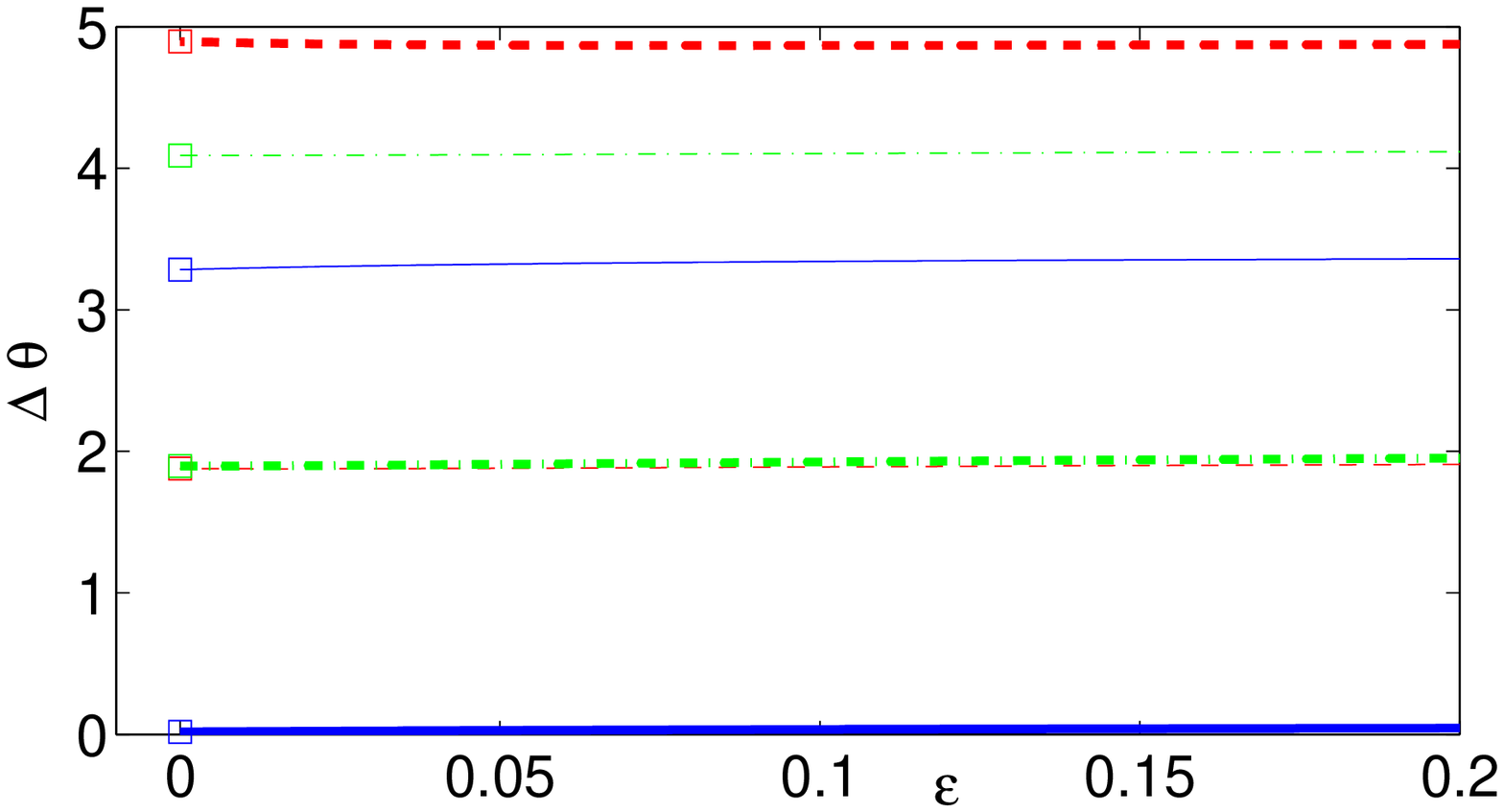}
\includegraphics[width=7cm,height=5cm,angle=0,clip]{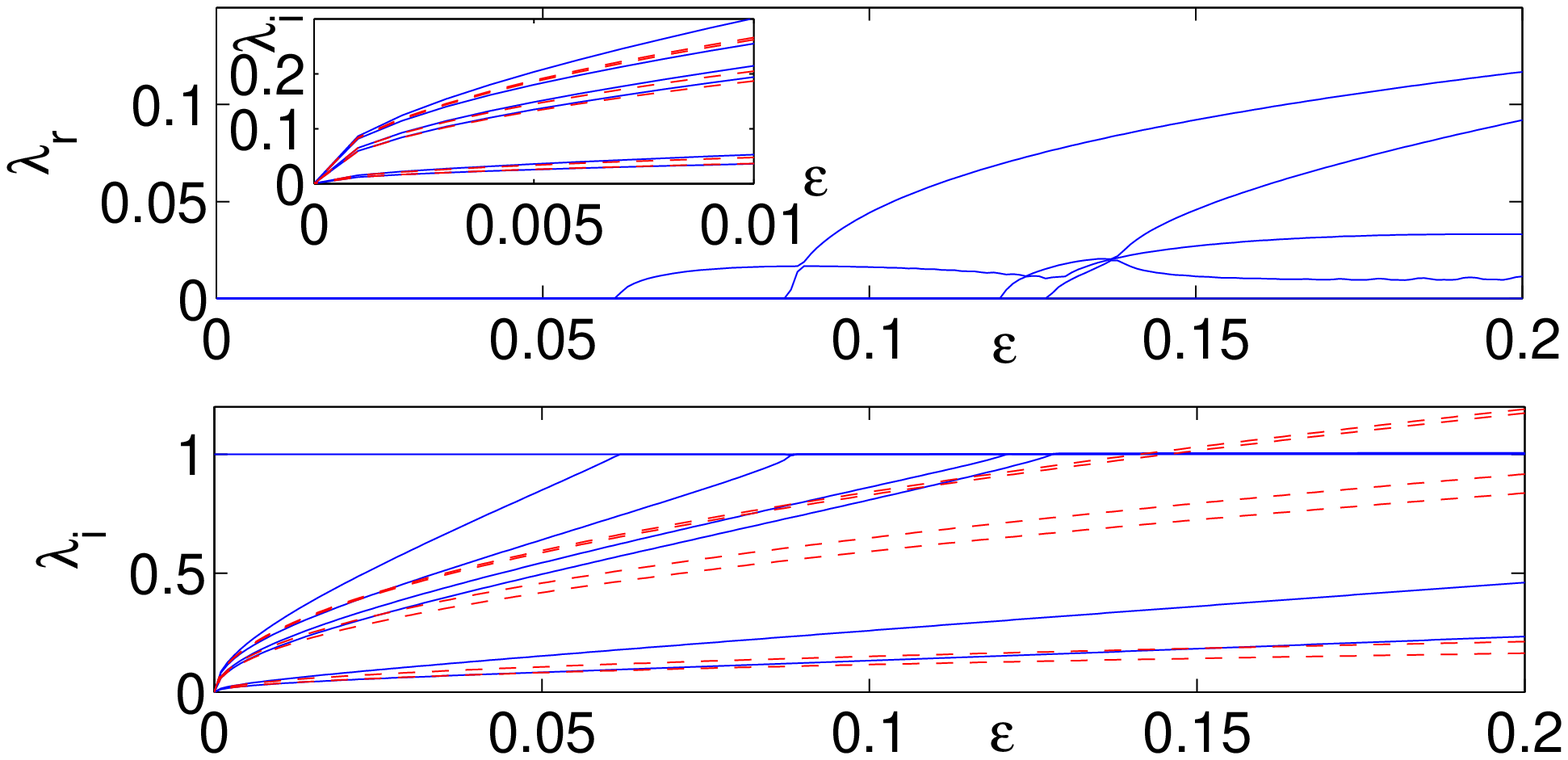}
\caption{A more complicated 1d example of a 7-site wave with nontrivial
phase distribution for $k_{n,n \pm 1}=k_{n,n \pm 2}=k_{n,n \pm 3}=1$. 
The top left panels
show the real and imaginary parts of the solution and its spectral plane
for $\varepsilon=0.2$; the top right shows the relative phases
$\Delta \theta_i=\theta_{i}-\theta_{1}$ for $i=2,\dots,7$ ($i=2$: blue
solid, $i=3$: red dashed, $i=4$: green dash-dotted, $i=5$ thick blue
solid, $i=6$: thick red dashed, $i=7$: thick green dash-dotted); the
theoretical anti-continuum limit is given by squares. Lastly, the 
bottom panels show the six relevant eigenvalues (blue solid lines)
and the corresponding theoretical predictions (red dashed lines).
The real (top) and imaginary (bottom) parts of the eigenvalues are
given -- it is clear that the solution becomes unstable for
$\varepsilon > 0.062$. This inset contains a blowup of the
imaginary parts for small 
$\varepsilon$ to showcase the agreement between numerics and theory.} 
\label{fig3}
\end{figure}

However, the above features such as the nontrivial phase
distribution are not a unique feature of cases with just NN
and NNN interactions. As an alternative example that shares
these features, we consider the case where 
$k_{nm}=\delta_{m,n \pm 1} + \delta_{m,n \pm 2} + \delta_{m,n \pm 3}$. 
Contrary to what might be naively expected, 
we do not find here the generalization of a vortex including 4-sites.
While the lowest order
persistence
conditions are satisfied by such a solution, higher order ones
are not. However, this does not preclude the existence of
genuinely complex solutions with a complicated phase structure.
An example of this kind involving $7$ sites is shown in Fig. 
\ref{fig3}. The phases of the 7 excited sites are found at the
AC limit to be  $\theta_1=0$ (since one of them can always be chosen
arbitrarily),
$\theta_2=3.2831$, $\theta_3=1.8780$, $\theta_4=4.0894$,
$\theta_5=0.01766$ and $\theta_7=1.8957$. Then, the corresponding
eigenvalue predictions from the reduced eigenvalue problem are:
$\lambda=\pm \sqrt{7.0792 \varepsilon} i$, 
$\lambda=\pm \sqrt{6.875 \varepsilon} i$, 
$\lambda=\pm \sqrt{4.2006 \varepsilon} i$,
$\lambda=\pm \sqrt{3.4994 \varepsilon} i$,
$\lambda=\pm \sqrt{0.2248 \varepsilon} i$,
$\lambda=\pm \sqrt{0.1334 \varepsilon} i$. These are compared
to the full numerical eigenvalues of the problem (as are
the existence results) in Fig. \ref{fig3}. Despite the 
complexity of the eigenvalue structure, the quality of
the agreement between theory and numerics is clear
from the eigenvalue inset and the phase comparisons (near the AC limit).
Notice that the solution becomes unstable for
$\varepsilon>0.062$ due to collision of the largest of the
above pairs with the continuous spectrum and increasingly
so due to additional collisions for 
$\varepsilon>0.088$, $\varepsilon>0.121$ and
$\varepsilon>0.128$.

\begin{figure}[tbp]
\includegraphics[width=8cm,height=5cm,angle=0,clip]{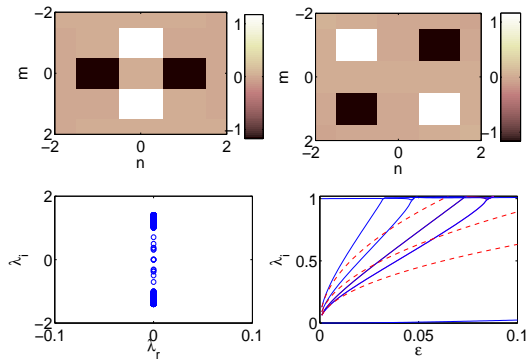}
\caption{An example of the stabilized $S=2$ vortex for 
$\varepsilon=0.02$ (the real contour of the solutions 
is shown in the top left and the imaginary
one in the top right). The spectral plane indicating the stability of
the solution is shown in the bottom left while the corresponding 
numerically obtained 
(imaginary) eigenvalues are in the bottom right (blue solid, as
compared with the red dashed theoretical prediction). 
The solution becomes unstable for $\varepsilon>0.032$.} 
\label{fig4}
\end{figure}

As our last example and in order to demonstrate a case where
the non-nearest-neighbor interactions have a direct impact
on stability, we consider a prototypical two-dimensional
case of an unstable structure with vorticity, namely the
vortex of topological charge $S=2$. It is well-known \cite{peli1d,old}
that this structure is unstable and a considerable effort has
been invested in stabilizing it either via defects \cite{pgk_djf}
or geometrically through a rhombic configuration \cite{joh2}. 
Here, we illustrate that NNN interactions {\it can also play
a stabilizing role} for this coherent structure. In particular,
we examined a special case (but verified that the result was
also robust for other choices), where $k_{{\bf n},{\bf m}}=1$ for
$|{\bf n}-{\bf m}|=1$,  $k_{{\bf n},{\bf m}}=1$ for
$|{\bf n}-{\bf m}|=2$, while  $k_{{\bf n},{\bf m}}=2$ for
$|{\bf n}-{\bf m}|=\sqrt{2}$ (where the last choice was
made because of the two lattice paths connecting the relevant
neighbors). In this case, (but also with other choices of
the relative weights), it is straightforward to show
that the square configuration with $\theta_j=j \pi/2$, where
$j=1,\dots,8$ remains a solution and that the corresponding
reduced eigenvalue problem has a single eigenvalue pair of
$\lambda=\pm 4 \sqrt{\varepsilon} i$, another single pair
of $\lambda=\pm \sqrt{8 \varepsilon} i$, and a quadruple
pair of $\lambda=\pm 2 \sqrt{\varepsilon} i$ (while one
pair is exactly at 0 and one is 0 to leading order but
also remains imaginary at higher orders). This stabilized
variant of the $S=2$ vortex can be observed in Fig. \ref{fig4},
where again the analytical predictions are compared to the
numerical results, yielding increasingly good agreement
once again for smaller values of $\varepsilon$. Notice
that the solution becomes destabilized for 
$\varepsilon>0.032$, while additional quartets arise for
$\varepsilon > 0.048$,
 $\varepsilon > 0.074$ and $\varepsilon > 0.087$.

\section{Conclusions}

In the present work, we have revisited the theme of
non-nearest-neighbor
interactions in settings of the discrete nonlinear
Schr{\"o}dinger type motivated by applications especially in optical
(but also in other areas such as atomic) physics. 
We illustrated that an analytical approach
enables a number of interesting conclusions and possibilities
that did not seem to have been realized before, to the best
of our knowledge, both as concerns the existence problem on the
infinite
lattice,
and as concerns the stability properties in the context
of localized solitary wave solutions.

In particular, we focused on the vicinity of the anti-continuum
limit of vanishing coupling. There, we could obtain explicit
conditions for the persistence of localized modes, as well as
analytical predictions for the eigenvalues of linearization around
such solutions. This analysis allows to conclude that the
non-nearest-neighbor setting has a number of special properties.
More specifically, even just next-nearest-neighbor interactions
(but also more general ones) create the possibility for
solutions with more complicated phase profiles than just
phases of $0$ and $\pi$ as was the case for nearest-neighbor
interactions \cite{peli1d}. 
On the other hand, these longer
range interactions can play a critical role in the stability of
solutions that were found to be unstable in the nearest-neighbor
setting. We presented a particular such example in the case of
the topological charge $S=2$ discrete two-dimensional vortex.

Although our work lays the ground for understanding some of
the main features of these non-nearest-neighbor settings,
a number of interesting questions are still outstanding.
In particular, it would be interesting if something could
be said more generally about the type of solutions (and phase
distributions) afforded
by interactions of different range/decay properties.
On the other hand, it would be interesting if a general classification
of stability could also be made for different types of intereaction
kernels, in higher or even in one dimension. These questions
are presently under investigation and will be reported
in future publications.

\vspace{5mm}

The warm hospitality of the Kirchoff Institute for Physics and
of the Institute for Physics of the University of Heidelberg (at
the last stage of this work),
as well as the support of NSF-DMS-0349023, NSF-DMS-0806762
and of the Alexander von Humboldt Foundation are gratefully acknowledged.

\end{document}